\documentclass[12pt]{article}
\usepackage{epsfig}
\usepackage{amssymb}
\usepackage{amsmath}
\usepackage{amsfonts}
\usepackage{graphicx}
\usepackage{mathrsfs}
\usepackage[dvips]{color}
\usepackage{multirow}

\newcommand{\R}{\mathbb{R}}
\newcommand{\C}{\mathbb{C}}

\newcommand{\fz}{\mathfrak{z}}

\newcommand{\bbe}{\mathbf{e}}

\newcommand{\bk}{\mathbf{k}}

\newcommand{\bE}{\mathbf{E}}

\newcommand{\be}{\begin{equation}}
\newcommand{\ee}{\end{equation}}
\newcommand{\bea}{\begin{eqnarray}}
\newcommand{\eea}{\end{eqnarray}}
\newcommand{\nn}{\nonumber}

\newcommand{\ed}{\end{document}}

\newcommand{\bi}{\begin{itemize}}
\newcommand{\ei}{\end{itemize}}

\newcommand{\bce}{\begin{center}}
\newcommand{\ece}{\end{center}}

\newcommand{\sE}{\mathscr{E}}

\newcommand{\RE}{{\rm Re}}

\begin{document}

\title{Blow-up solutions of Helmholtz equation for a Kerr slab with a complex linear and nonlinear permittivity}

\author{Varga Kalantarov$^1$, Ali Mostafazadeh$^{1,2,}$\thanks{E-mail address:
amostafazadeh@ku.edu.tr}, and Neslihan Oflaz$^{2}$, \\[6pt]
Departments of Mathematics$^1$ and Physics$^{1,2}$, Ko\c{c} University,\\  34450
Sar{\i}yer, Istanbul, Turkey}

\date{ }
\maketitle

\begin{abstract}

We show that the Helmholtz equation describing the propagation of transverse electric waves in a Kerr slab with a complex linear permittivity and a complex Kerr coefficient admits blow-up solutions provided that the real part of the Kerr coefficient is negative, i.e., the slab is defocusing.This result applies to homogeneous as well as inhomogeneous Kerr slabs whose linear permittivity and Kerr coefficient are continuous functions of the transverse coordinate. For an inhomogeneous Kerr slab, blow-up solutions exist if the real part of Kerr coefficient is bounded above by a negative number.


\end{abstract}

The emergence of blow-up solutions of nonlinear wave equations is a
well-known mathematical phenomenon
\cite{keller,kaplan,wong,glassey}. Recently it was shown that such
solutions can be realized in a genuine scattering setup and lead to
a nonlinear resonance effect and a corresponding nonlinear
amplification scheme for electromagnetic waves \cite{p147}. These
results make use of blow-up solutions of the Helmholtz equation for
transverse electric (TE) waves propagating in a homogeneous Kerr
slab with a real and negative Kerr coefficient. A valid question of
practical importance is whether blow-up solutions and the associated
resonance effect can be realized in a Kerr slab involving losses and
inhomogeneities. The purpose of the present article is to address
this question.

Because losses in a Kerr slab correspond to situations that the linear and nonlinear permittivity of the slab takes complex values, we consider a Kerr slab of thickness $L$ which has a complex Kerr coefficient $\sigma$ and a complex linear relative permittivity $\hat\varepsilon_l$. Suppose that we choose a Cartesian coordinate system $\{(x,y,z)\}$ in which the slab fills the volume confined between the planes $z=0$ and $z=L$, and $\sigma$ and $\hat\varepsilon_l$ do not depend on $x$ and $y$. Then the  propagation of time-Harmonic TE waves in this slab is described by the Helmholtz equation \cite{p147},
    \be
    \sE''(z)+k^2\left[\hat\varepsilon_l(z)-\sin^2\theta+\sigma(z)|\sE(z)|^2\right]\sE(z)=0,~~~~~~~~0\leq z\leq L,
    \label{H-eq}
    \ee
where $\sE(z)$ is the complex amplitude of the electric field, $\bE(z,t)=e^{-i\omega t}\sE(z)\hat\bbe_y$, $\omega:=ck$, $c$ is the speed of light in vacuum, $k$ is the wavenumber, $\hat\bbe_j$ is the unit vector along the $j$-axis, and $\theta$ is the incidence angle of the wave, so that its wavevector takes the form $\bk=k(\cos\theta\,\hat\bbe_z+\sin\theta\,\hat\bbe_x)$.

For a homogeneous Kerr slab with no loss or gain,  $\sigma$ and $\hat\varepsilon_l$ are real constants, and we can express the general solution of (\ref{H-eq}) in terms of Jacobi elliptic functions \cite{chen}. In particular when $\sigma<0$, i.e., for a defocusing lossless (and gainless) homogeneous Kerr slab, we can construct closed-form analytic blow-up solutions of (\ref{H-eq}). A simple example is
    \be
    \sE_\star(z):=A e^{i\varphi}
    \sec\left[\frac{\pi}{4}\left(\frac{z}{z_\star}+1\right)\right],
    \label{sol}
    \ee
    where
    \begin{align}
    &A:=\sqrt{\frac{2r}{-\sigma}},
    &&r:=\hat\varepsilon_l-\sin^2\theta,
    &&z_\star:=\frac{\pi}{4k\sqrt r},
    \label{z-star}
    \end{align}
   $\varphi$ is an arbitrary real number, and we have assumed that $\hat\varepsilon_l>\sin^2\theta$, \cite{p147}. Because $\sE_\star(z)$ diverges for $z=z_\star$, it is a blow-up solution of (\ref{H-eq}) provided that $L\geq z_\star$.

   We intend to show that the presence of losses does not obstruct the existence of blow-up solutions of (\ref{H-eq}) as long as the real part of $\sigma$ remains negative. This turns out to hold for the more general situations where the linear permittivity and the Kerr coefficient of the slab are continuous complex-valued functions of $z$. To state this result in a more compact and precise manner, we introduce:
    \begin{align}
    &\fz:=k z, &&\phi(\fz):=\sE(\fz/k), &&r(\fz):=\hat\varepsilon_l(\fz/k)-\sin^2\theta,
    &&s(\fz):=\sigma(\fz/k),
    \end{align}
and write the Helmholtz equation (\ref{H-eq}) as
    \be
    \phi''(\fz)+\left[r(\fz)+s(\fz)|\phi(\fz)|^2\right]\phi(\fz)=0.
    \label{eq}
    \ee
The following theorem is the main result of this article.
    \begin{itemize}
    \item[]{\bf Theorem~1:} Let $a,b\in\R$ and $r,s:\R\to\C$ be continuous functions such that for all $\fz\in\R$,
        \bea
        \RE[r(\fz)]\leq a,
        \label{condi-1}\\
        \RE[s(\fz)]\leq b < 0.
        \label{condi-2}
        \eea
Then there is a positive real number $\fz_\star$ and a solution $\phi_\star(\fz)$ of (\ref{eq}) in the interval $[0,z_\star)$ such that $\phi_\star(\fz)$ blows-up at $\fz=\fz_\star$.
    \end{itemize}
In view of the existence theorem for the solutions of ordinary differential equations \cite{hale}, (\ref{eq}) has a unique local solution for any given pair of initial conditions, $\phi(0)=c_0$ and $\phi'(0)=c_1$. To prove Theorem~1, it suffices to choose $c_0$ and $c_1$ such that the corresponding solution does not extend to $[0,\infty)$. To do this we make use of the following lemma that is proven in Ref.~\cite{glassey}.
    \begin{itemize}
    \item[]{\bf Lemma~1:} Let $u:[0,\infty)\to[0,\infty)$ and
    $h:[0,\infty)\to\R$ be functions such that $u$ is twice differentiable, the following conditions hold
        \bea
        &&\alpha:=u(0)>0,~~~~~~~\beta:=u'(0)>0,
        \label{e1}\\[6pt]
        &&h(s)\geq 0~\mbox{for all $s\geq \alpha$,}
        \label{e2}\\[6pt]
        &&h(u(\fz))\leq u''(\fz)~\mbox{for all $\fz\geq 0$},
        \label{e3}
        \eea
    and $\int_\alpha^s h(s')ds'$ exists for all $s>\alpha$. Then for every $\fz$ in the
    domain of $u$,  $u'(\fz)>0$  and
        \be
        \fz\leq \int_\alpha^{u(\fz)}\frac{ds}{\sqrt{\beta^2+2\int_\alpha^s h(s')ds'}}
        .
        \label{e4}
        \ee
    \end{itemize}

    \begin{itemize}
    \item[]{\bf Proof of Theorem~1:} Let $\phi(\fz)$ be the solution of (\ref{eq}) fulfilling initial conditions, $\phi(0)=c_0$ and $\phi'(0)=c_1$, for some $c_0,c_1\in\C\setminus\{0\}$, and $u:[0,\infty)\to[0,\infty)$ and $h:[0,\infty)\to[0,\infty)$ be functions defined by
        \bea
        u(\fz)&:=&\frac{|\phi(\fz)|^2}{2},
        \label{u}\\
        h(s)&:=&-2(a+2bs)s.
        \label{h}
        \eea
    Then it is easy to show that
        \begin{align}
        &u'=\RE(\phi^*\phi'),
        \label{e5}\\
        &u''=|\phi'|^2-2[\RE(r)+2\RE(s)u]u.
        \label{e6}
        \end{align}
    In view of (\ref{u}) and (\ref{e5}),
        \begin{align}
        &\alpha:=u(0)=|c_0|^2/2, &&\beta:=u'(0)=|c_0c_1|\cos(\vartheta_1-\vartheta_0),
        \label{e7}
        \end{align}
    where $\vartheta_0$ and $\vartheta_1$ are respectively the argument (phase angle) of $c_0$ and $c_1$, so that $c_0=|c_0|e^{i\vartheta_0}$ and $c_1=|c_1|e^{i\vartheta_1}$. According to (\ref{e7}), $\alpha>0$. Demanding that $\cos(\vartheta_1-\vartheta_0)>0$, we have $\beta>0$. Therefore (\ref{e1}) holds.    Next, choose $c_0$ such that $|c_0|>\sqrt{2a/|b|}$. Then $\alpha>a/|b|$, and because $b<0<\alpha$, for all $s>\alpha$, we have
    \[h(s)=2(-a+2|b|s)s>2(-a+2|b|\alpha)\alpha>2(-a+|b|\alpha)\alpha>0.\]
    Hence $h(s)$ fulfills (\ref{e2}).
    Furthermore,  (\ref{e6}) together with (\ref{condi-1}), (\ref{condi-2}), and (\ref{h}) imply
        \be
        u''(\fz)\geq -2\big\{\RE[r(\fz)]+2\RE[s(\fz)]u(\fz)\big\}u(\fz)\geq h(u(\fz)).
        \ee
    This establishes (\ref{e3}) and completes the proof that $u(\fz)$ and $h(s)$ fulfill the hypothesis of Lemma~1. According to this lemma, $u'(\fz)>0$ and (\ref{e4}) holds. If $\phi(x)$ did not have a singularity in  $[0,\infty)$, (\ref{e4}) would have held for all $\fz\in[0,\infty)$. This implies that for all $\fz\in[0,\infty)$,
        \be
    \fz\leq \int_\alpha^{u(\fz)}\frac{ds}{\sqrt{\beta^2+2\int_\alpha^s h(s')ds'}}<
    \int_\alpha^\infty\frac{ds}{\sqrt{\beta^2+2\int_\alpha^s h(s')ds'}}=:\gamma.
    \ee
   This a contradiction, because $\gamma$ is finite. In fact, evaluating $\int_\alpha^s h(s')ds'$ and using the result in the expression for $\gamma$, we have
    \bea
    \gamma&=&\int_\alpha^\infty\frac{ds}{\sqrt{p(s-\alpha)}}
    =\int_0^\infty\frac{dt}{\sqrt{p(t)}}\nn\\
    &<&\int_0^\infty\frac{dt}{\sqrt{q(t)}}
    =\frac{\Gamma(1/3)\Gamma(7/6)}{\sqrt{\pi}}
    \left(\frac{3}{\beta |b|}\right)^{1/3}<\frac{2.023}{(\beta |b|)^{1/3}},\nn
    \eea
 where
    \begin{align}
    &p(t):=\frac{8}{3}|b| t^3+4(2\alpha |b|-a)t^2+8\alpha (\alpha |b|-a) t+\beta^2,\nn\\
    &q(t):=\frac{8}{3}|b| t^3+\beta^2,\nn
    \end{align}
$\Gamma$ is the Euler's Gamma Function, and we have made use of the fact that $\alpha$ and $\alpha |b|-a$ are positive real numbers.~~~$\square$
        \end{itemize}

The above proof of Theorem~1 shows that whenever $c_0c_1\neq 0$ and $\cos(\vartheta_1-\vartheta_0)>0$, there is a real number $\fz_\star$ smaller than $2.023\times (\beta |b|)^{-1/3}$ such that the solution of (\ref{eq}) satisfying the initial conditions, $\phi(0)=c_0$ and $\phi'(0)=c_1$, blows up at $\fz=\fz_\star$. In terms of the physical parameters of the Kerr slab, this means that the Helmholtz equation (\ref{H-eq}) for the slab will have a blow-up solution, if the real part of its Kerr coefficient $\sigma$ is bounded above by a negative number $b$,
   $\sE(0)\sE'(0)\neq 0$, the phase angles for $\sE(0)$ and $\sE'(0)$, namely $\theta_0$ and $\theta_1$, satisfy $\cos(\vartheta_1-\vartheta_2)>0$, and
    \be
    L\geq L_\star:=\frac{2.023}{\left[ k^2 \big|b\,\sE(0)\sE'(0)\big| \cos(\vartheta_1-\vartheta_0)\right]^{1/3}}.
    \label{L-bound}
    \ee
For the solution (\ref{sol}), $L_\star=1.274/k\sqrt r=1.622\,
z_\star>z_\star$, where we have used (\ref{z-star}). Therefore
(\ref{L-bound}) is consistent with the fact that whenever $L\geq
z_\star$, (\ref{sol}) gives a blow-up solution of (\ref{H-eq}) for a
homogeneous Kerr slab with real linear permittivity and real and
negative Kerr coefficient.\vspace{6pt}

\noindent {\bf Acknowledgements}: This work has been supported by
the Scientific and Technological Research Council of Turkey
(T\"UB\.{I}TAK) in the framework of the project no: 114F357, and by
the Turkish Academy of Sciences (T\"UBA).

\ed
\begin{thebibliography}{99}


\bibitem{keller} J.~B.~Keller,  On solutions of nonlinear wave equations,
Comm.\ Pure Appl.\ Math.\ {\bf 10}, 523-530 (1957).

\bibitem{kaplan}  S.~Kaplan, On the growth of solutions of quasilinear parabolic
equations, Comm.\ Pure Appl.\ Math.\ {\bf 16}, 305-330 (1963).

\bibitem{wong} P.-K.~Wong, Bounds for solutions to a class of
nonlinear second-order differential equations, J.~Diff.\ Eq.~\textbf{7}, 139 (1970).

\bibitem{glassey} R.~T.~Glassey, Blow-up theorems for nonlinear wave equations, Math.~Z.\ {\bf 132}, 183 (1973).

\bibitem{p147} A.~Mostafazadeh, H.~Ghaemi-dizicheh, and S.~Hajizadeh, Blowing-up light, preprint arXiv: 1807.00644.

\bibitem{chen} W.~Chen and D.~L.~Mills, Optical response of a nonlinear dielectric film, Phys.\ Rev.~B {\bf 35}, 524 (1987) and optical behavior of a nonlinear thin film with oblique S-polarized incident wave, {\bf 38}, 12814 (1988).

\bibitem{hale} J.~K.~Hale, Ordinary Differential Equations, Dover, New York, 1997.



\end{thebibliography}
